# Cherenkov loss factor of short relativistic bunches: general approach

S.S. Baturin[1], A.D. Kanareykin[1,2]

[1]. St.Petersburg Electrotechnical University LETI, St.Petersburg Russia 197376

[2]. Euclid Techlabs LLC, Solon, OH 44139

**Abstract:**

The interaction of short relativistic charged particle bunches with waveguides and other accelerator system components is a critical issue for the development of X-ray FELs (free electron lasers) and linear collider projects. Wakefield Cherenkov losses of short bunches have been studied previously for resistive wall, disk-loaded, corrugated and dielectric loaded waveguides. It was noted in various publications [1] that if the slowdown layer is thin, the Cherenkov loss factor of a short bunch does not depend on the guiding system material and is a constant for any given transverse cross section dimensions of the waveguides. In this paper, we consider a new approach to the analysis of loss factors for relativistic short bunches and formulate a general integral relation that allows calculation of the loss factor for a short relativistic bunch passing an arbitrary waveguide system. The loss factors calculated by this new method for various types of waveguides with arbitrary thickness slowdown layers, including infinite media, are shown to be equivalent to those obtained by a conventional approach.

**Introduction.**

Relativistic, high intensity and small emittance electron bunches are the basis of linear collider (ILC, CLIC) and FEL (LCLS, NGLS, X-FEL etc.) projects among many others. These bunches excite Cherenkov wakefields as long as electrons pass through the accelerating structures or other longitudinally extended components of a beam line (pipes, collimators, bellows). The longitudinal size of these bunches are significantly smaller than the fundamental mode wavelength of the wakefields excited: this defines a "short" bunch in this context. For example, the longitudinal size of the 1 nC LCLS bunch are less than 20-30 $\mu m$, while the ILC linear collider project plans to operate with 300 $\mu m$ 3 nC bunches. The fundamental wakefield mode wavelength is defined by the transverse dimensions of the guiding system. For short bunches the ratio $\sigma_z / a$ of the bunch length $\sigma_z$ to the characteristic gap dimension $a$ has to be extremely small: for example, in the projects mentioned above this ratio is around 0.009 [1].

Additional interest in the short range wakefields generated by a relativistic beam arises from an approach that is used for reducing the energy spread of the beam [2-4,23]—a concept dubbed the ''wakefield silencer'' at the Argonne Wakefield Accelerator Facility [2,3], a beam "dechirper" [4]. Typically, in a linac-based X-ray FEL the beam is not fully compressed in the final chicane, leaving it with an energy chirp (a correlation between energy and position of the particles in the bunch) that needs to be removed. FELs that use lower-frequency SRF (superconducting RF) accelerating structures (XFEL, NGLS) generate wakefields that have a magnitude too small to be able to eliminate the chirp. A dielectric loaded waveguide [2,3] or a metallic beam pipe or rectangular waveguide with small, periodic corrugations [4] have been proposed as dedicated dechirpers. The dielectric "silencer" approach requires detailed loss factor analysis for a short relativistic bunch passing through a waveguide with arbitrary dielectric layer thickness [5].

Evidently, the longitudinal Cherenkov loss factor, or the limiting value of losses of a point-like unit charge passing the "stretched" beamline components, is of great interest for the projects referred above. The equivalence and exact matching of the loss factor of the beams passing through various types of waveguides with thin slowdown layers (features of the waveguide interior other than a smooth perfectly conducting surface) have been noted previously [1,12]. Indeed, the loss factor attains exactly the same value for all periodic, cylindrically symmetric structures [1], for a resistive pipe [6], a disk-loaded accelerator structure [7], a pipe with small periodic corrugations [8,9], and a cylindrical metal structure with a thin dielectric layer [10]. The same equivalence of the loss factor can be found for non-cylindrical structures as well [12]. For example, for a planar or rectangular all-metal waveguide with resistive walls [6,11], small corrugations [12,13] or a thin dielectric liner [8,10,14,15] the loss factor would be the same for structures with equal apertures: it is a constant that is dependent on transverse dimensions but independent of material properties [12].

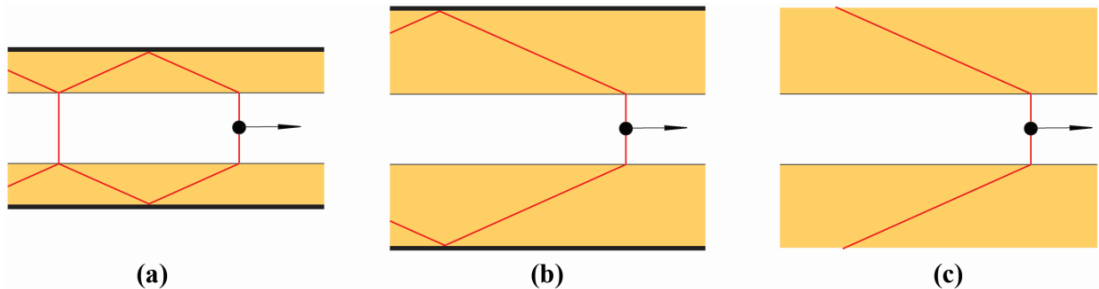

**Figure 1.** Cherenkov wakefield cones of a point-like charge moving along (a) a waveguide with a thin dielectric layer and a metal sleeve; (b) waveguide with a thick dielectric layer; (c) infinite medium.

The observation that loss factor calculations for a point-like charge in a waveguide of fixed transverse dimensions but different types of slowdown layers give equivalent results leads to the conclusion that a more general consideration is required. We have developed a new approach to the loss factor analysis of relativistic point-like charges under only assumption that the phase velocity of the Cherenkov radiation in the waveguide layers is less than the speed of light (the bunch is assumed relativistic, $V = c$). This allows the formulation of a general integral relation for the loss factor of a short relativistic bunch passing an arbitrary guiding system, independent of the waveguide shape (disk-loaded), the properties of the walls (dielectric or corrugated) or the waveguide material properties (dielectric constant or resistive walls). Moreover, a special conclusion of this approach is that the loss factor of the waveguides with slowdown layers or resistive materials does not depend on the layer thickness and gives the same results as those for the loss factor of the bunch passing the channel in an infinite dielectric or any other slowdown media. Note that the Cherenkov radiation of the bunch moving through an infinite slowdown medium are the same as for a particle passing along a channel inside unbounded dielectric if the channel transverse dimensions are close to the Cherenkov radiation wavelength [16]. That gives the equivalence of short bunch losses in a layered waveguide to the well-known formula for the Cherenkov loss factor of a point-like particle traversing an infinite dielectric medium [17].

**Cherenkov wakefields and loss factors.**

Field-particle interactions in high energy physics are usually described in terms of wake and impedance formalism: more details can be found elsewhere [6,18]. Consider a point charge moving with the speed of light along the axis of a vacuum accelerating structure. A test charge also moving with the speed $v=c$ at a distance $s$ behind this point-like bunch will experience fields of the first charge if the bunch separation $s$ is greater than the so-called "catch-up" distance [1,18,19]. To describe the interaction between the first and second particle a function $W(s)$ called the wake potential was introduced [18,19]. Vanishing of the wake functions everywhere in front of a relativistic particle is a consequence of causality: the wake potential is equal to zero for $s<0$. Wake potentials can be expressed with the eigenmode decomposition, where $w_n(s)$ - is the n$^{th}$ mode wake function:

$$W(s) = \sum_n \kappa_n w_n(s) \quad , \quad \kappa = W(0) = \sum_n \kappa_n . \tag{1}$$

Here $\kappa_n$ is the eigenmode loss factor. The total loss factor $\kappa$ is usually defined as (1).

It was shown that in the case of a thin corrugation layer [1,4,6-13] or dielectric [4,6,8-10] the total loss factor is equal to the loss factor of the fundamental mode of the structure and expressed by:

$$\kappa_c = \frac{Z_0 c}{2\pi a_c^2} . \tag{2}$$

The loss factor of a conductive cylindrical pipe, Fig.2a, can be found elsewhere [1,6,18-19] and is also equal to (2), where $Z_0 = 377\,\Omega$ is the impedance of vacuum, $c$ is the speed of light, $a_c$ is the pipe radius.

Meanwhile, the loss factor of a relativistic point-like charge passing through a planar structure, Fig.2c can be expressed as:

$$\kappa_p = \frac{Z_0 c}{2\pi a_p^2} \frac{\pi^2}{16} \tag{3}$$

[1,12] where $a_p$ is the vacuum half gap. The fact that total loss factor does not depend on the material or corrugation parameters was noticed and discussed previously [1,12]. Using a simple integration over the structure cross section presented in the next section we can prove that formulas (2) and (3) can be applied in the case of a thick layer and thus show that the material thickness and properties do not affect the total loss factor. Moreover, the calculation method is quite simple and is based on the analog of Gauss's law. We do not include for the moment frequency dispersion.

**Integral transformation.**

Let us consider circulation of the magnetic field on the boundary of a metal and apply Green's theorem

$$\oint \mathbf{H} \cdot d\mathbf{l} = \iint_{S_\perp} (\nabla \times \mathbf{H}) \cdot d\mathbf{S}, \tag{4}$$

where the integral on the right hand side is calculated over the cross section $S_\perp$ of the waveguide, and the left hand side integral is taken along the metal sleeve of the waveguide. We write the integral of the right hand side of (4) in the scalar form, using Ampère's law:

$$\nabla \times \mathbf{H} = \frac{\partial \mathbf{D}}{c \partial t} - \frac{en\mathbf{V}}{c}; \qquad \iint_{S_\perp}(\nabla \times \mathbf{H}) \cdot d\mathbf{S} = \iint_{S_\perp}\left(\frac{\partial D_z}{c \partial t} - \frac{enV}{c}\right) dS. \tag{5}$$

Here $n = 4\pi N \delta(z - Vt)\delta(x - x_0)\delta(y - y_0)$ is the particle density, where N is the number of electrons. Substituting $\zeta = z - Vt$ and rewriting (4) with (5) becomes:

$$-\oint \mathbf{H} \cdot d\mathbf{l} = \beta \iint_{S_\perp} \frac{\partial D_z}{\partial \zeta} dS + 4\pi q \beta \delta(\zeta). \tag{6}$$

Now consider the waveguide cross sectional area $S_\perp^q$ that includes the charge. If we assume that our particle is moving with the speed of light $V \approx c$, then we can also assume that the longitudinal self-Coulomb field $E_z^{coul} \approx 0$. Using the fact that in the medium the phase speed of light is lower than the speed of the moving charge, we can conclude that in the cross section $S_\perp^q$ the nonzero field is localized only in the vacuum gap. From this position in the limit $V \to c$ one can rewrite (6) for $S_\perp^q$ as:

$$\iint_{S_\perp^q} \frac{\partial D_z}{\partial \zeta} dS = -4\pi q \delta(\zeta) - \int_{l_1} \mathbf{H} \cdot d\mathbf{l}. \tag{7}$$

Here the integral on the right side is calculated along the metal surfaces of the waveguide $l_1$ that are not covered by the material, and $D_z$ is the z component of the electric displacement vector that corresponds to the Cherenkov field. In the case where there are not uncovered metal walls one can see that as long as the nonzero field is localized only in the vacuum gap, the integral on the right side correspondingly is equal to zero and (7) simplifies to:

$$\iint_{S_\perp^q} \frac{\partial D_z}{\partial \zeta} dS = -4\pi q \delta(\zeta). \tag{8}$$

Now decompose the integral on the left side into an integral over the vacuum channel $S_V$ and an integral over $S_D$, the cross section of the medium:

$$\iint_{S_V} \frac{\partial E_z^V}{\partial \zeta} dS = -4\pi q \delta(\zeta) + \varepsilon \iint_{S_D} \frac{\partial E_z^D}{\partial \zeta} dS . \qquad (9)$$

Integration of (9) with respect to $\zeta$, with $E_z^V(-\infty)=0$ and $E_z^D(-\infty)=0$ leads to

$$\iint_{S_V} E_z^V(\zeta)dS = -4\pi q \int_{-\infty}^{\zeta} \frac{d\theta(x)}{dx} dx + \varepsilon \iint_{S_D} E_z^D(\zeta)dS . \qquad (10)$$

Here $\theta(x)$ is the Heaviside theta-function. If now we set $\zeta=0$ because the flux through $S_D$ is zero it immediately gives

$$\iint_{S_V} E_z^V(0)dS = -2\pi q , \qquad (11)$$

and in the case $\zeta=0^+$ a factor of 2 has to be applied. Formula (11) gives a simple connection between the longitudinal electric field in the cross section of the bunch and the total bunch charge, which looks like a classical Gauss's law. Using this expression we will show that radiation losses and transverse distribution of the electric field could be found using the well-known technique of conformal mapping.

**Cylindrical waveguide, longitudinal loss factor**

Consider a round cross-section of the metal waveguide with an arbitrary slowdown layer along the metal walls and a vacuum channel along the axis, Fig. 2a.

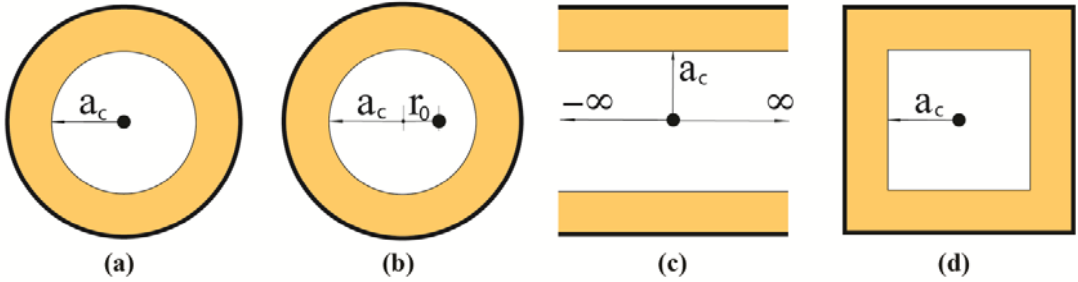

**Figure 2.** Cross section of the considered metal waveguides with the slowdown layers (yellow): (a) cylindrical; (b) cylindrical with displaced charge; (c) planar; (d) square.

It is easy to show that in the case of a cylindrical structure and $V \to c$ $E_z^V(\zeta)$ does not depend on transverse coordinates when the particle is moving along the $z$-axis of a cylinder. Thus for a cylinder from (11) in SI units we have:

$$E_z^c(0) = -\frac{Z_0 cq}{2\pi a_c^2}, \qquad E_z^c(0^+) = -\frac{Z_0 cq}{\pi a_c^2} . \qquad (12)$$

Here $a_c$ is the radius of the vacuum gap. It should be noticed that no assumptions on the thickness of the slowdown (dielectric, corrugation etc.) were used while obtaining expressions

(11) and (12). Typically only thin layer approximation is considered for the loss factor analysis [1,4,8,9,12]. The proposed general integral formalism is based only on the assumption that (a) the bunch is a ultra-relativistic one and (b) the speed of the bunch is exceeding the speed of light in the layer. One can conclude that the expressions (11) and (12) are true for any layer thickness including boundaryless media, Fig.1c.

**Cylindrical waveguide. Kick factor.**

If the beam traverses the waveguide off-axis, Fig.2b, a deflecting field will affect the beam [1,6,12]; if the off-set distance of the beam is relatively small, only the dipole mode of almost the same frequency as the fundamental mode will be excited. If the beam gets deflected with the considerable off-set, additional mulipoles will contribute to the dipole deflection force [6,10].

The deflecting force factor or the "kick factor" for the cylindrical waveguide with the arbitrary slowdown layer will be presented in this section. Let's consider a conformal transformation of a circle $|\omega| \leq a_c$ on a circle $|z| \leq a_c$ such as that the point $r_0$ ( $Arg[r_0]=0$ ) of a first circle transforms into the center $z=0$ of a second one. The correspondent mapping is given then by:

$$z = a_c^2 \frac{\omega - r_0}{a_c^2 - \omega r_0}. \tag{13}$$

We consider now an integral over the vacuum gap cross-section along the $z$ - plane and rewrite it for the $\omega$ -plane:

$$\int E_z^c dS_z = \int E_z^c J dS_\omega. \tag{14}$$

Here $J$ is the determinant of the Jacobi matrix, $dS_z$ - is the elementary square of the z -plane $dS_\omega$ - is the surface square element on the $\omega$ -plane. Using the fact that conformal transformation is an analytical function one can write

$$J = \left|\frac{dF(\omega)}{d\omega}\right|^2 = \left|\frac{d}{d\omega} a_c^2 \frac{\omega - r_0}{a_c^2 - \omega r_0}\right|^2. \tag{15}$$

Assuming $|\omega| = r$ and $Arg[\omega] = \varphi$ one can obtain

$$J(r,\varphi) = \frac{a_c^4 (a_c^2 - r_0^2)^2}{(a_c^4 + r^2 r_0^2 - 2 r r_0 a_c^2 \cos(\varphi))^2}. \tag{16}$$

As long as $E_z^c$ is a constant we can conclude from (14) that field distribution $E_z^{dp}$ over the $\omega$ - plane can be found as $E_z^{dp}(r,r_0,\varphi) = J(r,\varphi) E_z^c$. Thus in the origin ( $r = r_0, \varphi = 0, \zeta = 0$ ) longitudinal field could be written as:

$$E_z^{dp}(r_0,r_0,0) = -\frac{2q}{a_c^2} \frac{1}{\left(1-\left(r_0/a_c\right)^2\right)^2}. \tag{17}$$

The radial part of Lorentz force could be calculated using Panofsky-Wenzel theorem [22]. The force derivative at the origin, also known as the kick factor, could be found if $r = r_0$ and $\varphi = 0$. Using (17) we obtain in SI units:

$$\kappa_\perp^c = \frac{1}{q^2 r_0} \frac{\partial F_r(r,0)}{\partial \zeta}\bigg|_{\substack{r=r_0 \\ \zeta=0}} = \frac{1}{q r_0} \frac{\partial E_z(r,0)}{\partial r}\bigg|_{\substack{r=r_0 \\ \zeta=0}} = \frac{Z_0 c}{2\pi a_c^4} \frac{4}{\left(1-(r_0/a_c)^2\right)^3}, \quad (18)$$

$$\kappa_\perp^c \approx \frac{2 Z_0 c}{\pi a_c^4}(1 + 3(r_0/a_c)^2); \quad r_0/a_c \ll 1. \quad (19)$$

One can see that for the small offsets $r_0/a_c \ll 1$ the first term of the loss factor (19) is equal to the well-known result for the loss factor of the pipe with small corrugation or resistive walls as expected [1,2,6,8,9]. Note the divergence of (18) at $r_0/a_c \to 1$ that corresponds to the dispersion-less model of the slowdown layer media. If the beam traverses along the layer the dispersion properties of the layer are critical for the Cherenkov losses and transverse deflecting force simulations and in reality the high frequency part of the spectrum is always limited [17]. The same kick factor divergence is observed with the mode decomposition simulations [10].

**Planar waveguide. Longitudinal loss factor.**

Let's consider a point-like charge moving along the symmetry axis in between two infinitely long plates, Fig.2c. The conformal transformation of a strip $\text{Re}[\omega] \in [-\infty, \infty]$, $\text{Im}[\omega] \in [0_c, 2a_c]$ onto the interior of a circle of radius $a_c$ $|z| \leq a_c$ can be written as

$$\frac{z}{a_c} = \tanh\left[\frac{\pi}{4}\left(\frac{\omega}{a_c} - i\right)\right]. \quad (20)$$

Following the same formalism as in (14)-(16) we calculate the Jacobi determinant of the mapping (20). Assuming $\text{Re}[\omega] = x$ and $\text{Im}[\omega] = y$ we have:

$$J(x,y) = \frac{\pi^2}{4}\left(\cosh\left[\frac{\pi x}{2a_c}\right] + \sin\left[\frac{\pi y}{2a_c}\right]\right)^{-2}. \quad (21)$$

From (14) for the center of the planar structure $x = 0, y = a_c$ with (20) one can obtain formulas of the longitudinal field at the point-like charge in SI units as:

$$E_z^p(0) = -\frac{Z_0 c q}{2 a_c^2} \frac{\pi}{16}; \quad E_z^p(0^+) = -\frac{Z_0 c q}{a_c^2} \frac{\pi}{16}. \quad (22)$$

**Square waveguide. Longitudinal loss factor.**

The loss factor for a square cross-section metal structure with a slowdown layer, Fig.2d, can be obtained using the Kristoffel-Schwarz integral that gives a conformal mapping of the inner part of a circle $|z|<a_c$ to a square with each side equal to $2a_c$:

$$\omega = \sqrt{\frac{2}{\pi}} \frac{\Gamma\left(\frac{3}{4}\right)}{\Gamma\left(\frac{5}{4}\right)} \int_0^z \frac{dt}{\sqrt{1-t^4}}, \qquad (23)$$

where $\Gamma(x)$ is the Euler Gamma function. Taking into account that $z = r\exp(i\varphi)$ one can write determinant of a Jacobi matrix for transformation of an elementary square as:

$$J(r,\varphi) = \frac{2}{\pi} \left[\frac{\Gamma\left(\frac{3}{4}\right)}{\Gamma\left(\frac{5}{4}\right)}\right]^2 \frac{1}{\sqrt{1+(r/a_c)^8 - 2(r/a_c)^4 \cos(4\varphi)}}, \quad \text{where} \quad \frac{2}{\pi}\left[\frac{\Gamma\left(\frac{3}{4}\right)}{\Gamma\left(\frac{5}{4}\right)}\right]^2 = \frac{1}{f} = 1.16. \quad (24)$$

Taking into account (14) we have

$$E_z^{rec}(r,\varphi) = \frac{E_z^{cyl}}{J(r,\varphi)}, \qquad E_z^{rec}(0,\varphi) = \frac{E_z^{cyl}}{J(0,\varphi)} = fE_z^{cyl}, \qquad (25)$$

where the right part of (25) is taken at the position of a point-like charge ($r = 0$); and we have

$$E_z^{rec}(0) = -f\frac{Z_0 cq}{2\pi a_c^2}; \quad E_z^{rec}(0^+) = -f\frac{Z_0 cq}{\pi a_c^2}; \quad f = 0.86. \qquad (26)$$

It should be noticed that we considered the coefficient $f$ is equal 0.86 but it can be calculated with any required accuracy using its analytical definition (24).

**Discussion of the results.**

Using (12), (22) and (26) one can obtain an expression for the loss factors of the point-like charges for the cylindrical $\kappa_c$, planar $\kappa_p$ and square $\kappa_p$ metal waveguides with any kind of slowdown layers along the metal walls. Corresponding formulas are presented in SI-units in Table 1 including the kick factor (19) for the cylindrical waveguide $\kappa_\perp^c$. Meanwhile the wake potentials at $\zeta = 0^+$ differ from the loss factors with a factor of 2 [18,19].

Formulas (2), (3) for the cylindrical and planar waveguides and those of Table 1 look identical, but we emphasize here that (2), (3) as derived in references [1,6,18,19] used assumptions on the the particular slowdown mechanism (corrugation, dielectric) while our formulas presented in

Table 1 were obtained in the general case for any type of material and slowdown layer thickness. The kick factor formula (19), Table 1, was obtained for the full solution including all multipoles, not only dipoles. Also, the loss factor formula, (26) Table 1, was derived for the waveguide with the square cross-section metal wall all covered with the slowdown layer. This formula can be used for dielectric wakefield acceleration or THz generation devices [3,15]. In addition to the resistance and roughness, the waveguide wall may have an oxide layer, which is usually a dielectric. This effect for the very short bunches was previously studied in a round pipe only [8,24].

**Table 1.** Longitudinal loss factors for various cross sections of the waveguides, Fig. 2 (a,c,d), and the kick for cylindrical waveguides (small offsets), Fig 2(b).

| cylindrical | planar | square | cylindrical; kick factor |
|---|---|---|---|
| $\kappa_c = \dfrac{Z_0 c}{2\pi a_c^2}$ | $\kappa_p = \dfrac{Z_0 c}{2\pi a_c^2} \dfrac{\pi^2}{16}$ | $\kappa_{sq} = 0.86 \dfrac{Z_0 c}{2\pi a_c^2}$ | $\kappa_\perp^c = \dfrac{2 Z_0 c}{\pi a_c^4}(1 + 3(r_0/a_c)^2)$ |

Loss factor determination often becomes a complicated problem and involves massive numerical mode summations. With the new approach shown above, one can use relatively simple and yet powerful tools for the calculation of the total loss factor. Using the integral relation on the basis of the cylindrical slowdown waveguide model the full loss factor of the structure can be calculated. It was shown that for other cross-section geometries one can obtain the loss factor by using a conformal mapping that allows finding the ratio of the loss factor for a cylindrical structure to that of the other structure of interest. It should be noted that the loss factor in this case is simply the value of the Jacobi matrix determinant at the origin, and the Jacobi determinant away from the origin gives the transverse structure of the loss factor.

**Summary.**

We considered the Cherenkov loss factor of a point-like electron bunch passing through cylindrical and planar waveguides lined with arbitrary slowdown layers. It was shown that the Cherenkov loss factor of the short bunch does not depend on the waveguide system material and is a constant for any given transverse dimensions and cross-sections of the waveguides. The equivalence and exact matching of the loss factor of the beams passing through various types of waveguides is analyzed. The same equivalence of the loss factor was found for planar structures as well. The kick factor formula known for the first term only was obtained as a full solution with all the multipoles included. Also at first the loss factor for the waveguide with the square cross-section metal wall covered with the slowdown layer was obtained. It was shown that with the proposed approach one can use a relatively simple method for the calculation of the total loss factor using an integral relation based on the cylindrical slowdown waveguide model. For other types of cross-section geometries (rectangular or elliptical) one can obtain the loss factor by using a conformal mapping from the solution for the cylindrical case that will be presented with the follow-up publication.

**Acknowledgements**.

Euclid Techlabs LLC acknowledges support from the DOE SBIR Program, Grant # DE-SC0006299. The authors are grateful to A. Zholents, K. Bane, G. Stupakov and P. Schoessow for useful discussions and suggestions.